\documentclass[11pt,twoside]{article}

\usepackage{asp2006}

\begin{document}
\title{What are the Luminous Compact Blue Galaxies?}   %%% Fill in title
\author{D.J. Pisano\altaffilmark{1}, C.A. Garland\altaffilmark{2}, 
R. Guzm\'an,\altaffilmark{3} J. P\'erez Gallego\altaffilmark{3}, 
F.J. Castander\altaffilmark{4}, N. Gruel\altaffilmark{3}}   

\altaffiltext{1}{NRAO, P.O. Box 2, Green Bank, WV 24944 USA; dpisano@nrao.edu}
\altaffiltext{2}{Natural Sciences Dept., Jeffords Science Center, Castleton 
State College, Castleton, VT 05735}
\altaffiltext{3}{Dept. of Astronomy, University of Florida, 211 Bryant Space 
Science Center, P.O. Box 112055, Gainesville, FL 32611}
\altaffiltext{4}{Institut de Ci\'encies de l'Espai, Campus UAB, 08193 
Bellaterra, Barcelona, Spain}

\begin{abstract} %%% Abstract to run on from here.
Luminous Compact Blue Galaxies (LCBGs) are common at z$\sim$1, contributing
significantly to the total star formation rate density.  By z$\sim$0, they
are a factor of ten rarer.  While we know that LCBGs evolve rapidly, we
do not know what drives their evolution nor into what types of galaxies they 
evolve.  We present the results of a single-dish HI survey of local LCBGs 
undertaken to address these questions.  Our results indicate that LCBGs have 
M$_{HI}$ and M$_{DYN}$ consistent with low-mass spirals, but typically exhaust 
their gas reservoirs in less than 2 Gyr.  Overall, the properties of LCBGs are 
consistent with them evolving into high-mass dwarf elliptical or dwarf 
irregular galaxies or low-mass, late-type spiral galaxies.  
\end{abstract}

\section{Introduction}
Luminous Compact Blue Galaxies (LCBGs) are a morphologically and 
spectroscopically diverse class of galaxies characterized by high luminosity 
(M$_B$~$<-$18.5), blue color (B$-$V$~<~$0.6) and high surface brightness 
(SB$_e <$ 21 mag arcsec$^{-2}$).  At z$\sim$1, LCBGs are common, contributing
45\% of the total star formation rate density, but they evolve rapidly being
ten times rarer by z$\sim$0.  As such, it is important to understand the 
role of LCBGs in galaxy formation and evolution.  Yet we do not know the 
basic nature of these galaxies nor their evolutionary fate.  They may 
evolve into dwarf ellipticals \citep{koo94}, irregular or low-mass spiral 
galaxies \citep{phillips97,guzman98}, or bulges of grand-design spirals 
\citep{hammer01}.  In order to discriminate between these possibilities,
we have begun a multi-wavelength study of the rare, local (D$\le$200~Mpc) 
LCBGs.  Our sample was selected from the Sloan Digital Sky Survey to match 
the rest-frame optical properties as LCBGs observed at z$\sim$1.  A key 
component of this study is the measurement of the HI flux and linewidths, 
allowing determinations of gas content, dynamical masses, and gas depletion 
timescales. Such measurements constrain the evolutionary possibilities for 
this class of galaxies. Background and preliminary results have been 
previously reported by \citet{garland04,garland05,garland07}; in this 
proceedings we present the first results from our expanded study of 142 LCBGs. 

\section{Observations}
Our single-dish HI data come 
from a number of sources:  57 LCBGs were observed with Arecibo,
the Green Bank Telescope (GBT), or Nan\c{c}ay, down to a 5$\sigma$~M$_{HI}$ 
detection limit of 2.5$\times$10$^8$M$_{\sun}$.  Data on the remaining 85 
galaxies come from \citet{meyer04}, \citet{giovanelli05,giovanelli07}, 
\citet{springob05}, and the GBT HI 
survey\footnote{http://www.cv.nrao.edu/\~{}rfisher/GalaxySurvey/galaxy\_survey.html}. 
We reanalyzed all of the spectra; only two galaxies were marginal 
detections and two were undetected.  

\section{Conclusions}

Our survey found that local LCBGs have a diverse range of HI properties, but 
that they are generally gas-rich with M$_{HI}\sim 10^{9-10} M_{\sun}$.  They 
have M$_{DYN}$ within R$_{25}$ of 10$^{10-11} M_{\sun}$.  These properties, 
as well as their mass-to-light and M$_{HI}$/M$_{DYN}$, are similar to
those of late-type spiral galaxies \citep{roberts94}.  LCBGs have specific 
star formation rates (SFR/M$_{DYN}$) similar to HII galaxies, and gas 
depletion timescales $<$2~Gyr for the majority of the population.  The sizes 
and HI linewidths of local LCBGs, which should not evolve dramatically, 
suggest that they will become massive dwarf ellipticals or dwarf irregulars, 
or late-type spiral galaxies.  If the z$\sim$1 LCBGs have similar gas 
properties, then we can 
infer that those LCBGs will follow similarly diverse evolutionary paths.  
This work confirms the results of \citet{garland04} with a much larger sample.
We have begun a program of mapping a number of these
galaxies in HI to search for gas-rich companions, signatures of outflows and
recent interactions, as well as to obtain resolved measurements of their 
rotation curves to derive their dynamical masses.  Finally, we are obtaining 
additional data ranging from the radio to the X-ray regime, to further 
understand the nature of these galaxies.

%\acknowledgements %%% Text of acknowledgements runs on after this command.

%%% THE BIBLIOGRAPHY
%%%
%%% CONSULT SECTION 3 OF "INSTRUCTIONS FOR AUTHORS" FOR HOW TO USE NATBIB.
%%% AUTHORS ARE ENCOURAGED TO USE EITHER THE "THEBIBLIOGRAPY" ENVIRONMENT
%%% BY UNCOMMENTING (DELETING THE "%" SYMBOL) THE COMMANDS BELOW, OR BY
%%% USING THE BIBTEX ENVIRONMENT. TO FIND OUT WHICH IS APPLICABLE TO YOUR
%%% CONTRIBUTION, CONSULT THE VOLUME EDITORS FOR YOUR PROCEEDINGS.
%%%

\end{document}